\documentclass[runningheads]{llncs}
\usepackage[T1]{fontenc}
\usepackage{graphicx}
\usepackage{color, wrapfig}
\usepackage{tabularx}
\usepackage{multirow}
\usepackage{caption}
\usepackage{subcaption}
\usepackage{amsmath}
\usepackage{float}
\usepackage{marvosym}

\newcommand{\repeatthanks}{\textsuperscript{\thefootnote}}
\begin{document}
\title{A Study on Accuracy, Miscalibration, and Popularity Bias in Recommendations}
\titlerunning{Accuracy, Calibration, and Popularity Bias in Recommendations}

\author{
Dominik Kowald$^{\textrm{(\Letter)}}$\inst{1,2}\thanks{Both authors contributed equally to this work.} \and
Gregor Mayr\inst{2}\repeatthanks \and
Markus Schedl\inst{3,4} \and
Elisabeth Lex\inst{2}}
\institute{Know-Center GmbH, Graz, Austria \\\email{dkowald@know-center.at} \and
Graz University of Technology, Graz, Austria \\\email{gregor.mayr@student.tugraz.at\\{dominik.kowald,elisabeth.lex}@tugraz.at} \and
Johannes Kepler University \& Linz Institute of Technology, Linz, Austria \\\email{markus.schedl@jku.at}
}

\authorrunning{Dominik Kowald, Gregor Mayr, Markus Schedl, and Elisabeth Lex}

\maketitle    

\begin{abstract}
Recent research has suggested different metrics to measure the inconsistency of recommendation performance, including the accuracy difference between user groups, miscalibration, and popularity lift. However, a study that relates miscalibration and popularity lift to recommendation accuracy across different user groups is still missing. Additionally, it is unclear if particular genres contribute to the emergence of inconsistency in recommendation performance across user groups. In this paper, we present an analysis of these three aspects of five well-known recommendation algorithms for user groups that differ in their preference for popular content. Additionally, we study how different genres affect the inconsistency of recommendation performance, and how this is aligned with the popularity of the genres. Using data from Last.fm, MovieLens, and MyAnimeList, we present two key findings. First, we find that users with little interest in popular content receive the worst recommendation accuracy and that this is aligned with miscalibration and popularity lift. Second, our experiments show that particular genres contribute to a different extent to the inconsistency of recommendation performance, especially in terms of miscalibration in the case of the MyAnimeList dataset.

\keywords{Recommender systems \and Popularity bias \and Miscalibration \and Accuracy \and Recommendation inconsistency \and Popularity lift}
\end{abstract}
\section{Introduction}

Recommender systems benefit users by providing personalized suggestions of content such as movies or music. 
However, we also know from previous research that recommender systems suffer from an inconsistency in recommendation performance across different user groups~\cite{abdollahpouri2019impact,ekstrand2018all}. 
One example of this inconsistency is the varying recommendation accuracy across different user groups, which could lead to unfair treatment of users whose preferences are not in the mainstream of a community~\cite{kowald2020unfairness,kowald2021support}. 
Other examples are inconsistencies between the input data of a recommender system and the recommendations generated, which could lead to recommendations that are either too popular and/or do not match the interests of specific user groups~\cite{abdollahpouri2019impact,ekstrand2018all}. Thus, popularity bias can be seen as one particular example of recommendation inconsistencies.

Apart from measuring recommendation accuracy differences across different user groups, related research~\cite{abdollahpouri2019impact} suggests quantifying the inconsistency of recommendation performance along two metrics, namely miscalibration and popularity lift. Miscalibration quantifies the deviation of a genre spectrum between user profiles and actual recommendations~\cite{steck2018,lin2020}. For example, if a user listens to songs belonging to 45\% pop, 35\% rock, and 20\% rap, whereas a calibrated recommendation list should contain the same genre distribution. 

Related research also proposes the popularity lift metric to investigate to what extent recommendation algorithms amplify inconsistency in terms of popularity bias~\cite{abdollahpouri2019unfairness,abdollahpouri2020connection}. This popularity lift metric quantifies the disproportionate amount of recommendations of more popular items in a system. For example, a positive popularity lift indicates that the items recommended are on average more popular than the ones in the user profile. Therefore, in the remainder of this paper, we refer to popularity lift as a metric that measures the popularity bias of recommendation algorithms. 

However, a study that relates miscalibration and popularity lift to recommendation accuracy across different user groups is still missing. 
We believe that the outcomes of such a study could help choose the most suitable recommendation debiasing methods for each user group. 
Additionally, it is unclear if particular genres contribute to the emergence of inconsistency in recommendation performance across user groups. This knowledge could be helpful, e.g., for enhancing recommendation debiasing methods based on calibration.

\vspace{2mm} \noindent \textbf{The present work.} 
In this paper, we contribute with a study on accuracy, miscalibration, and popularity bias of five well-known recommendation algorithms that predict the preference of users for items, i.e., UserItemAvg, UserKNN, UserKNNAvg~\cite{hug2020surprise}, NMF~\cite{luo2014efficient}, and Co-Clustering~\cite{george2005scalable} in the domains of music (Last.fm), movies (MovieLens), and animes (MyAnimeList). We split the users in each dataset into three user groups based on the low, medium, and high inclination towards popular content, which we call LowPop, MedPop, and HighPop, respectively.
With this, we aim to shed light on the connection between accuracy, miscalibration, and popularity bias in recommendations. 

Furthermore, in this paper, we investigate what genres in the user groups are particularly affecting recommendation inconsistency across the algorithms and domains.
With this, we aim to understand if particular genres contribute to the emergence of inconsistency in recommendation performance, and if this is aligned with the popularity of the genres.  

\vspace{2mm} \noindent \textbf{Findings and contributions.} 
We find that LowPop consumers consistently receive the lowest recommendation accuracy, and in all investigated datasets, miscalibration is the highest for this user group. In terms of popularity lift, we observe that all algorithms amplify popularity bias. 

Concerning our analysis on the level of genres, we find that there are indeed genres that highly contribute to inconsistency, especially in terms of miscalibration in the case of the MyAnimeList dataset. In sum, the contributions of our paper are four-fold:
\begin{enumerate}
    \item We extend three well-known datasets from the field of recommender systems with genre information to study the inconsistency of recommendation performance.
    \item We evaluate five well-known recommendation algorithms for accuracy, miscalibration, and popularity lift.
    \item We inspect recommendation inconsistency on the genre level and show that different genres contribute differently to the emergence of inconsistency in recommendation performance.
    \item To foster the reproducibility of our work, we share the extended datasets and source code used in our study with the research community.
\end{enumerate}

\section{Related Work}
Bias in information retrieval and recommender systems is an emerging research trait, and related  works have shown multiple ways to quantify different biases in a system~\cite{baeza2020bias,lesota2021analyzing}. One such bias is the popularity bias, which arises due to items with higher popularity getting recommended more often than items with lower popularity. Works~\cite{ekstrand2018all} have found, that not all users are affected identically, with some user groups receiving more inconsistent recommendations than others. Ekstrand et al.~\cite{ekstrand2018all,kowald2022ecir}, for example, found inconsistencies in recommendation accuracy among demographic groups, with groups differing in gender and age showing statistically significant differences in effectiveness in multiple datasets. The authors evaluated different   recommendation algorithms and identified varying degrees of utility effects.

Abdollahpouri et al.~\cite{abdollahpouri2019unfairness,abdollahpouri2020connection,abdollahpouri2019impact} also contributed to this line of research and introduced two metrics to quantify the inconsistency in recommendation performance from the user’s perspective. The first one is the miscalibration metric, which quantifies the misalignment between the genre spectrum found in a user profile and the genre spectrum found in this user's recommendations. 
The second one is the popularity lift metric, which measures to what extent a user is affected by popularity bias, i.e., the unequal distribution of popular items in a user profile and this user's recommendations. In datasets from the movie domain, they found that users that are more affected by popularity bias also receive more miscalibrated results. Similarly, Kowald et al.~\cite{kowald2020unfairness} analyzed popularity bias and accuracy differences across user groups in the music domain. 
The authors found that the popularity lift metric provided different results in the music domain than in the movie domain due to repeat consumption patterns prevalent in the music-listening behavior of users.

In this paper, we extend these works by connecting miscalibration and popularity lift to recommendation accuracy across different user groups. Additionally, we examine if particular genres contribute to the emergence of recommendation inconsistency across user groups and datasets. 
With this, we hope to inform research on popularity bias mitigation methods. As an example,~\cite{abdollahpouri2021user} has proposed in-processing methods for debiasing recommendations based on calibration. We believe that our findings on which genres contribute to miscalibrated results could be used to enhance these methods. Additionally, related research has proposed post-processing methods to re-rank recommendation lists~\cite{abdollahpouri2019managing,adomavicius2011improving}. We believe that our findings for the connection of accuracy and popularity lift for different user groups could help choose the right users for whom such re-ranking should be performed.

\section{Method}
In this section, we describe the datasets, the experimental setup, and the evaluation metrics used in our study.

\begin{table}[!t]
\centering
\caption{Dataset statistics including the number of users $|U|$, items $|I|$, ratings $|R|$, and distinct genres $|C|$ as well as sparsity and rating range $R$-range.}
\resizebox{\textwidth}{!}{
\begin{tabular}{l|rrrr|rrr|l}
\hline
Dataset       & $|U|$  & $|I|$ & $|R|$ & $|C|$ & $|R|/|U|$ & $|R|/|I|$ & Sparsity & $R$-range \\ \hline
LFM        & $3,\!000$ & $131,\!188$ & $1,\!417,\!791$ & $20$ & 473 &  $11$ & $0.996$ & $[1-1,000]$ \\
ML     & $3,\!000$ &   $3,667$ &   $675,\!610$ & $18$ & 225 & $184$ & $0.938$ & $[1-5]$ \\
MAL   & $3,\!000$ &   $9,450$ &   $649,\!814$ & $44$ & 216 &  $69$ & $0.977$ & $[1-10]$ \\ \hline
\end{tabular}}
\label{tab:datasets}
\end{table}

\subsection{Datasets}
\begin{figure}[!t]
	\centering
 \resizebox{\textwidth}{!}{
	\begin{subfigure}[b]{\textwidth}
		\centering
		\includegraphics[width=\textwidth]{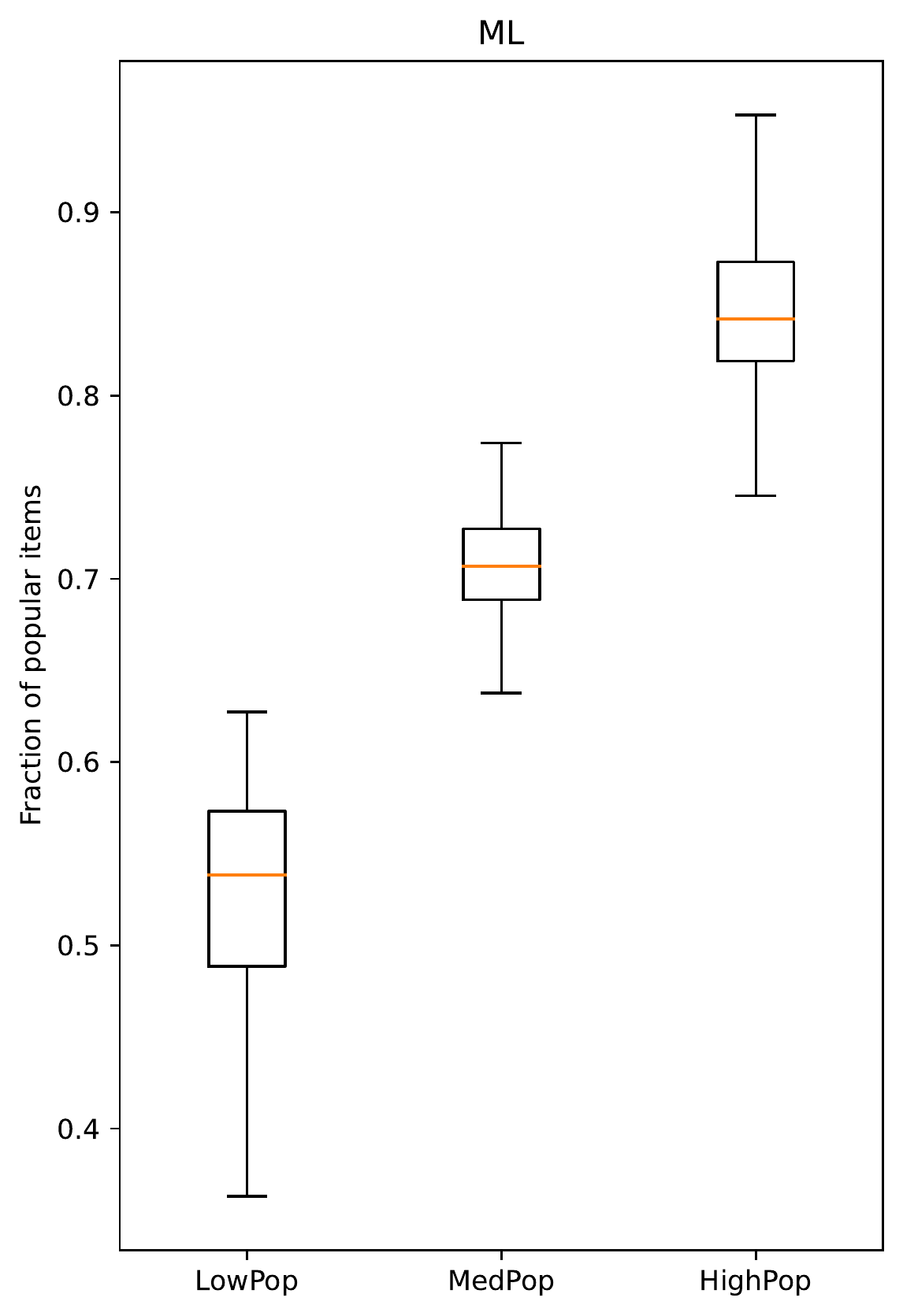}
	\end{subfigure}
    \begin{subfigure}[b]{\textwidth}
    \hspace{1mm}
		\centering
		\includegraphics[width=\textwidth]{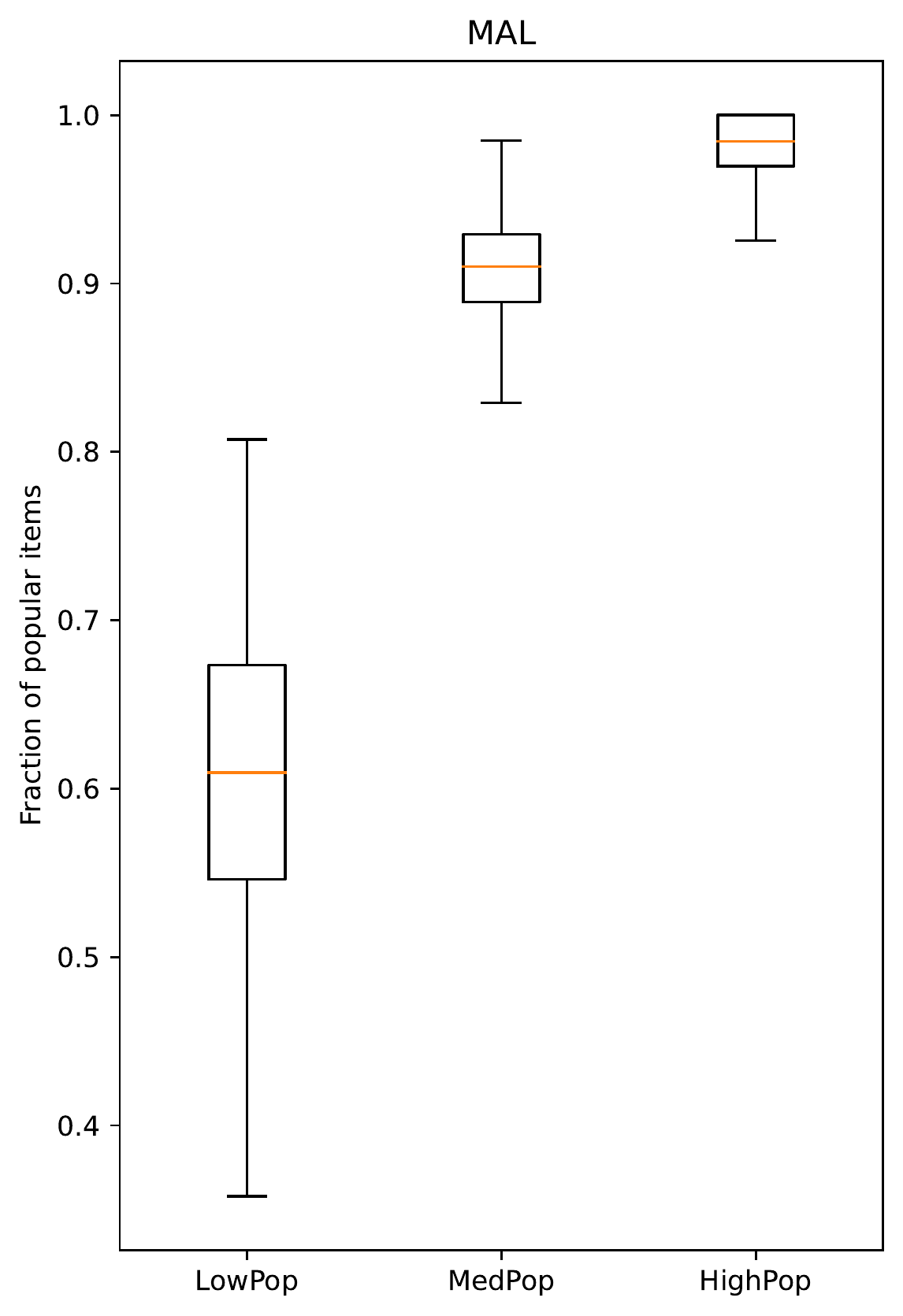}
	\end{subfigure}
    \begin{subfigure}[b]{\textwidth}
    \hspace{1mm}
		\centering
		\includegraphics[width=\textwidth]{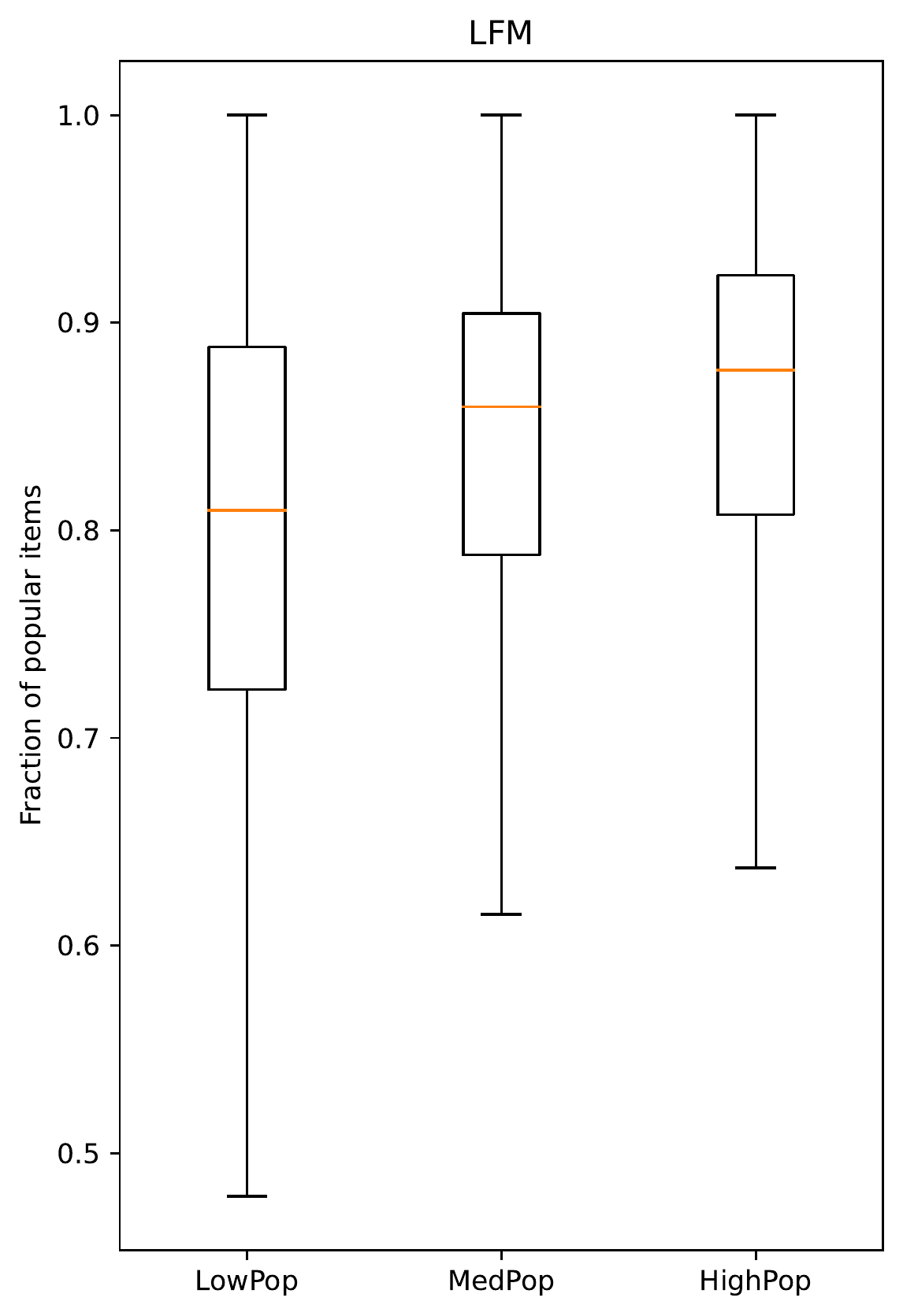}
	\end{subfigure}
	\hfill
 }
	\caption{Boxplots depicting the fraction of popular items in the user profiles for the three user groups and datasets. The LowPop group has the smallest ratio of popular items, compared to MedPop and HighPop. In the LFM dataset, this difference is not as apparent as in the other datasets, due to repeat consumption patterns in music listening behavior.}
	\label{fig:boxplots}
\end{figure}

We use three different datasets in the domains of music, movies, and animes. Specifically, we use dataset samples from Last.fm (LFM), MovieLens (ML), and MyAnimeList (MAL) provided in our previous work~\cite{kowald2022ecir}\footnote{We do not use the BookCrossing dataset due to the lack of genre information.}. Here, each dataset consists of exactly 3,000 users, which are split into three equally-sized groups with 1,000 users each. We use 1,000 users per user group to be comparable with previous works that also used groups of this size. 
The groups are created based on the users' inclination toward popular items. Following the definitions given in~\cite{kowald2022ecir}, we define a user $u$'s inclination towards popular items as the fraction of popular items in $u$'s user profile. We define an item $i$ as popular if it is within the top-20\% of item popularity scores, i.e., the relative number of users who have interacted with $i$. We term the group with the lowest, medium, and highest inclination toward popular items \textit{LowPop}, \textit{MedPop}, and \textit{HighPop}, respectively. In Figure~\ref{fig:boxplots}, we show boxplots of the fraction of popular items in the user profiles of the three groups for our three datasets.

In Figure~\ref{fig:boxplots}, we show boxplots depicting the fraction of popular items in the user profiles for the three user groups and datasets. We see that the LowPop user group has the smallest ratio of popular items, compared to MedPop and HighPop. In the case of the LFM dataset, this difference is not as apparent as in the case of the other datasets, due to repeat consumption patterns in music listening behavior.

Basic statistics of the datasets can be found in Table~\ref{tab:datasets}, and we share our dataset samples via Zenodo\footnote{\url{https://doi.org/10.5281/zenodo.7428435}}. In the following, we give more details on these datasets and how we extend them with genre information. Additionally, we analyze the popularity distributions in the datasets on the levels of ratings and users to give context for our study on genre level, which follows later on.  

\vspace{2mm} \noindent \textbf{Last.fm (LFM).} 
The LFM dataset sample used in our study is based on the LFM-1b dataset~\cite{schedl2016lfm} and the subset used in \cite{kowald2022ecir}. It contains listening records from the music streaming platform Last.fm. 
We include only listening records to music artists that contain genre information. Genre is acquired by indexing Last.fm's user-generated tags (assigned to artists) with 
the 20 main genres from the AllMusic database (top-3: rock, alternative, pop).  
When comparing the LFM dataset sample in Table~\ref{tab:datasets} with the one from~\cite{kowald2022ecir}, we notice that the number of artists $|I|$ decreases from 352,805 to 131,188, which means that there is no genre information available in LFM for a large set of the long-tail artists. However, in terms of ratings, this leads to a relatively small reduction in ratings from 1,755,361 to 1,417,791. 
Following our previous work~\cite{kowald2022ecir}, we interpret the number of times a user has listened to an artist as a rating score, scaled to a range of [1; 1,000] using min-max normalization. We perform the normalization on the level of the individual user to ensure that all users share the same rating range, in which the user's most listened artist has a rating score of 1,000 and the user's least listened artist has a rating score of 1.

\vspace{2mm} \noindent \textbf{MovieLens (ML).} 
Our ML dataset sample is based on the ML-1M dataset provided by the University of Minnesota~\cite{harper2015theMD}. Here, we gather the genre information for movies directly from the original dataset\footnote{\url{https://grouplens.org/datasets/movielens/1m/}}, which provides genres for all movies and contains 18 distinct genres (top-3: comedy, drama, action). 
With respect to sparsity, ML is our densest dataset sample, while LFM is our sparsest one. 

\vspace{2mm} \noindent \textbf{MyAnimeList (MAL).} 
The MAL dataset used in our study is based on a recommender systems challenge dataset provided by Kaggle. As in the case of ML, the original dataset\footnote{\url{https://www.kaggle.com/CooperUnion/anime-recommendations-database}} already provides genre information for each item, which leads to 44 distinct genres (top-3: comedy, action, romance). However, one special characteristic of MAL is that this dataset also contains implicit feedback (i.e., when a user bookmarks an anime). Following~\cite{kowald2022ecir}, we set the implicit feedback to an explicit rating of 5. 
In terms of the number of ratings, MAL is the smallest dataset used in our study, while LFM is the largest one.

\begin{figure}[!t]
	\centering
\resizebox{\textwidth}{!}{
	\begin{subfigure}[b]{\textwidth}
		\centering
		\includegraphics[width=0.85\textwidth]{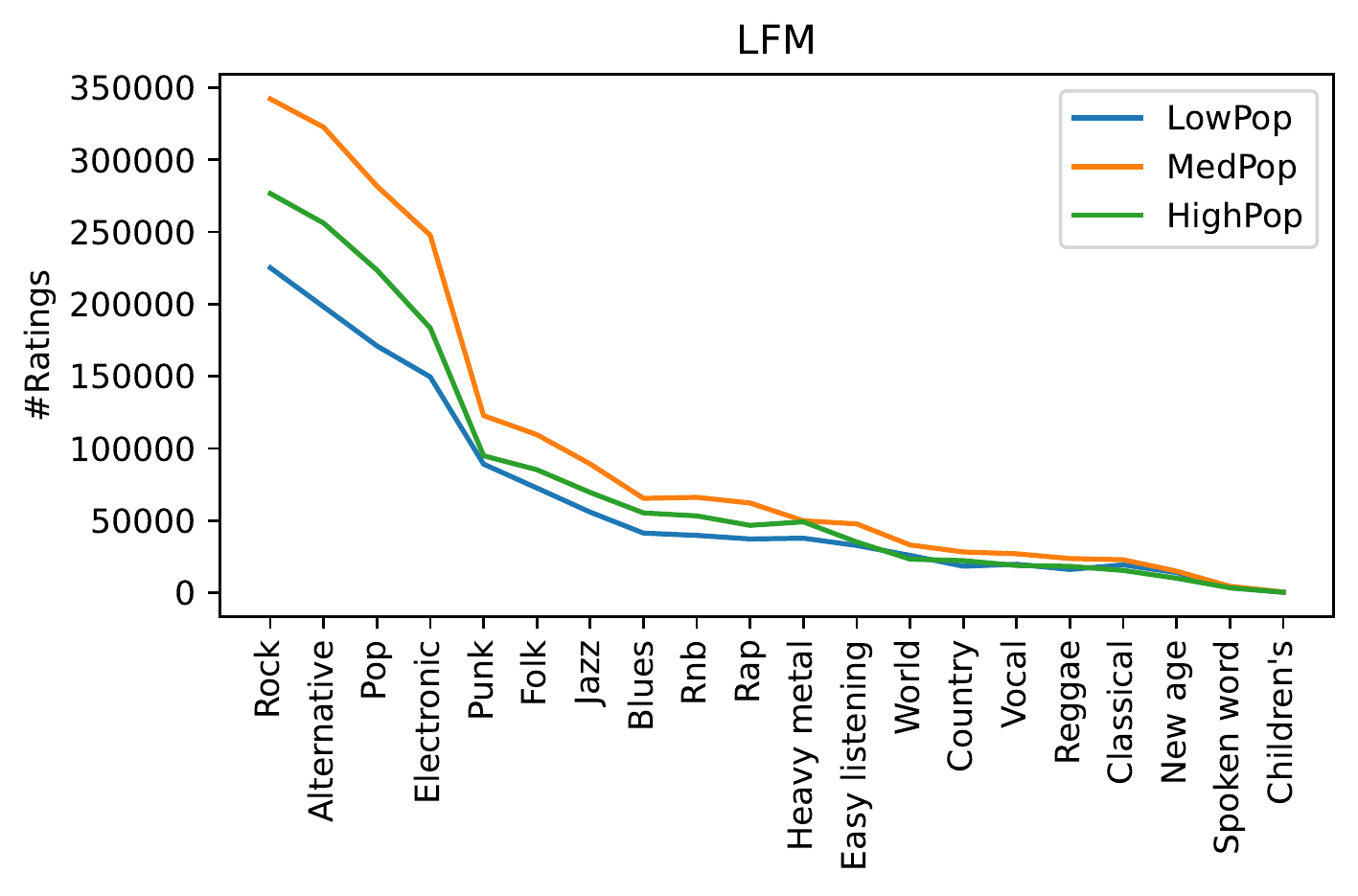}
	\end{subfigure}
	\begin{subfigure}[b]{\textwidth}
 \hspace{-2mm}
		\centering
		\includegraphics[width=0.85\textwidth]{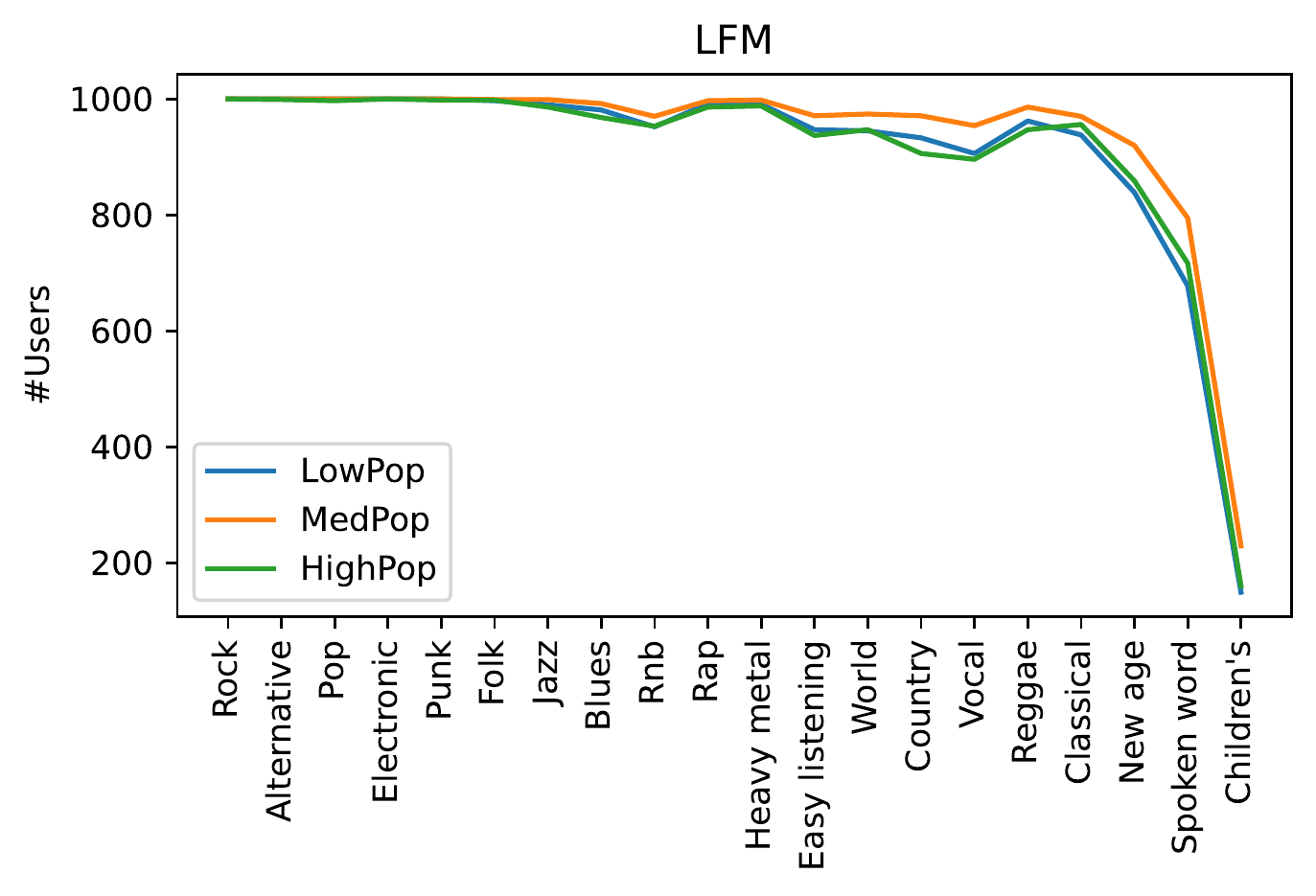}
	\end{subfigure}
	\hfill
 }
\resizebox{\textwidth}{!}{
	\begin{subfigure}[b]{\textwidth}
 \hspace{-2mm}
		\centering
		\includegraphics[width=0.86\textwidth]{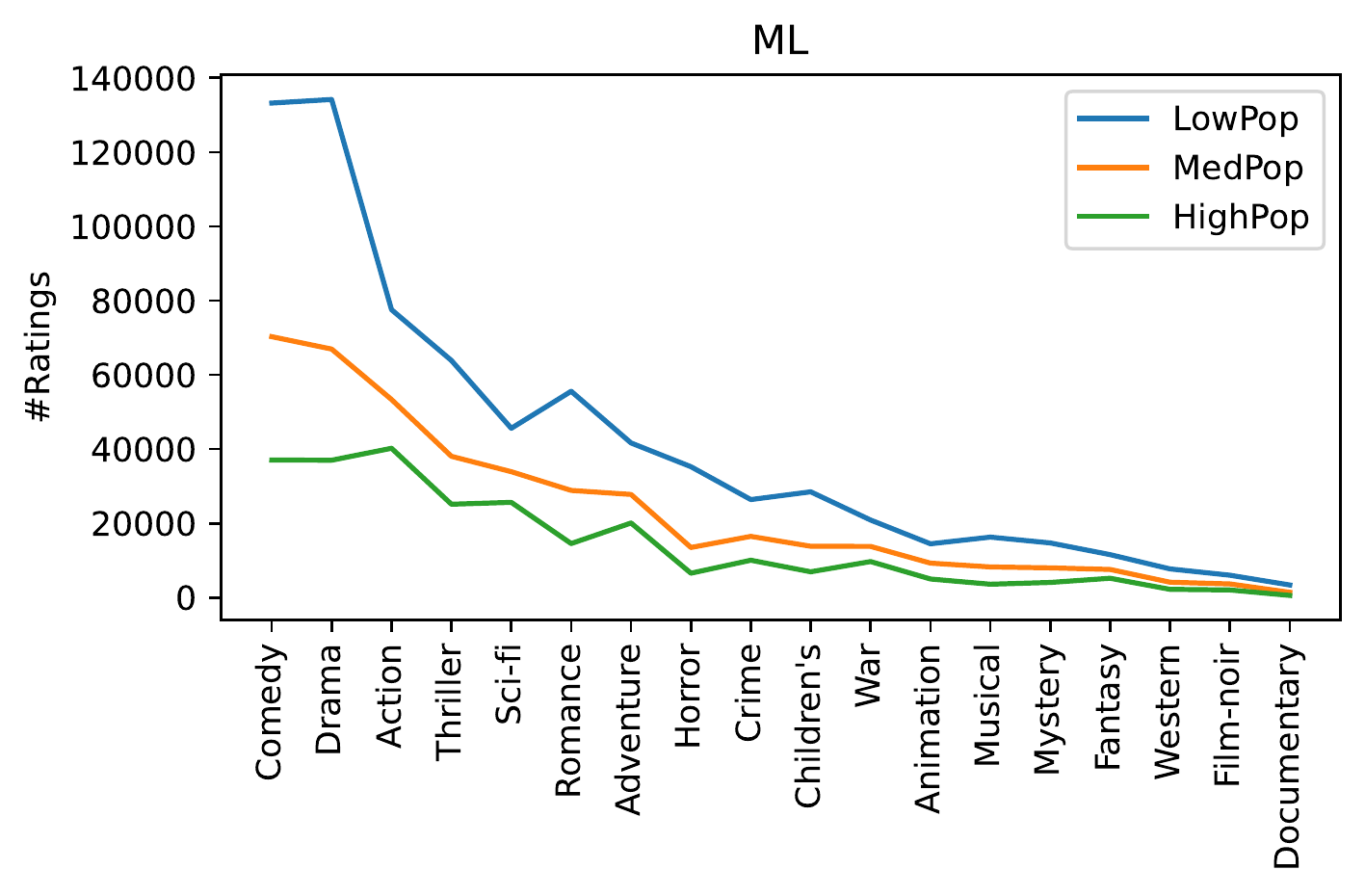}
	\end{subfigure}
	\begin{subfigure}[b]{\textwidth}
		\centering
		\includegraphics[width=0.85\textwidth]{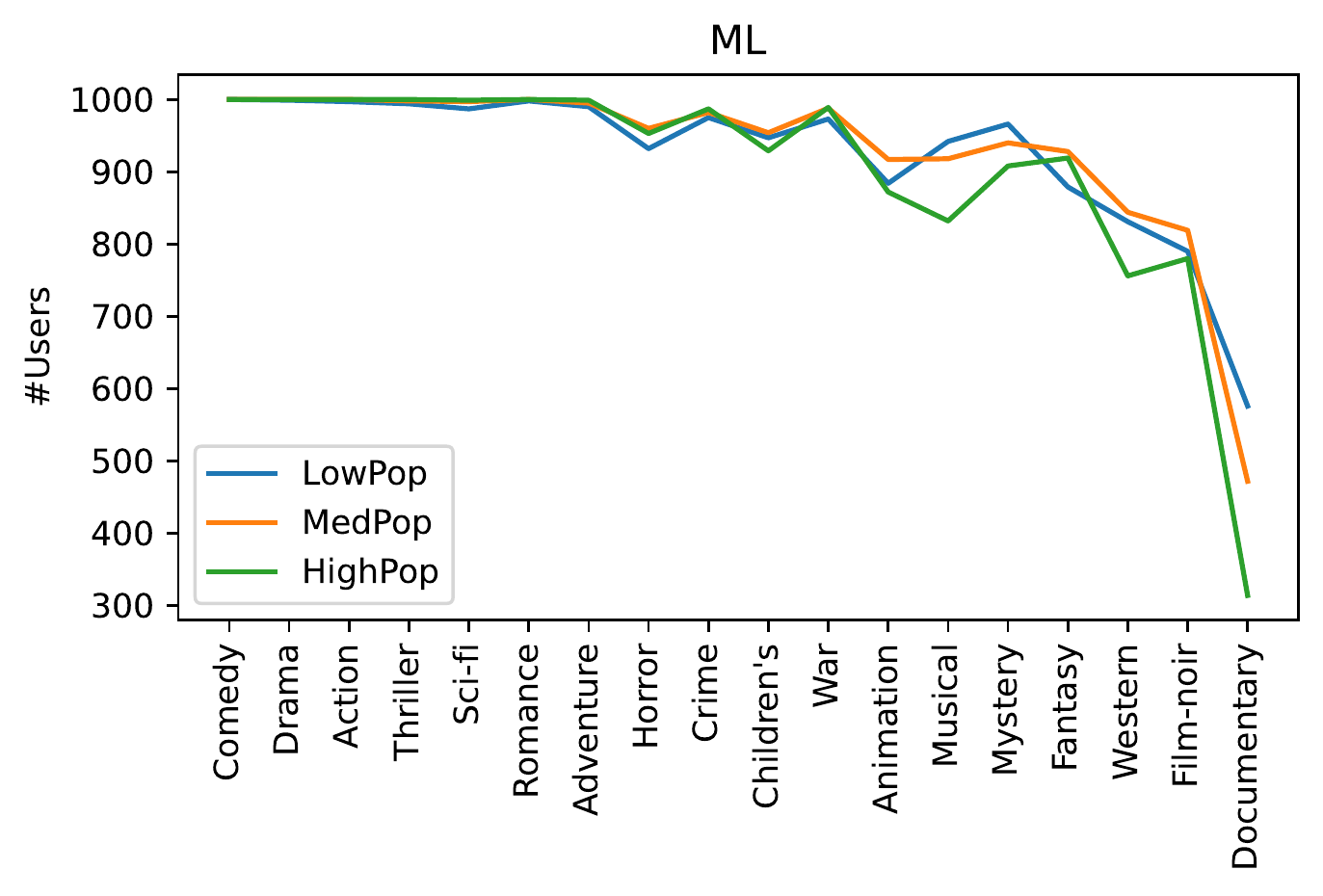}
	\end{subfigure}
	\hfill
 }
 \resizebox{\textwidth}{!}{
	\begin{subfigure}[b]{\textwidth}
		\centering
		\includegraphics[width=0.85\textwidth]{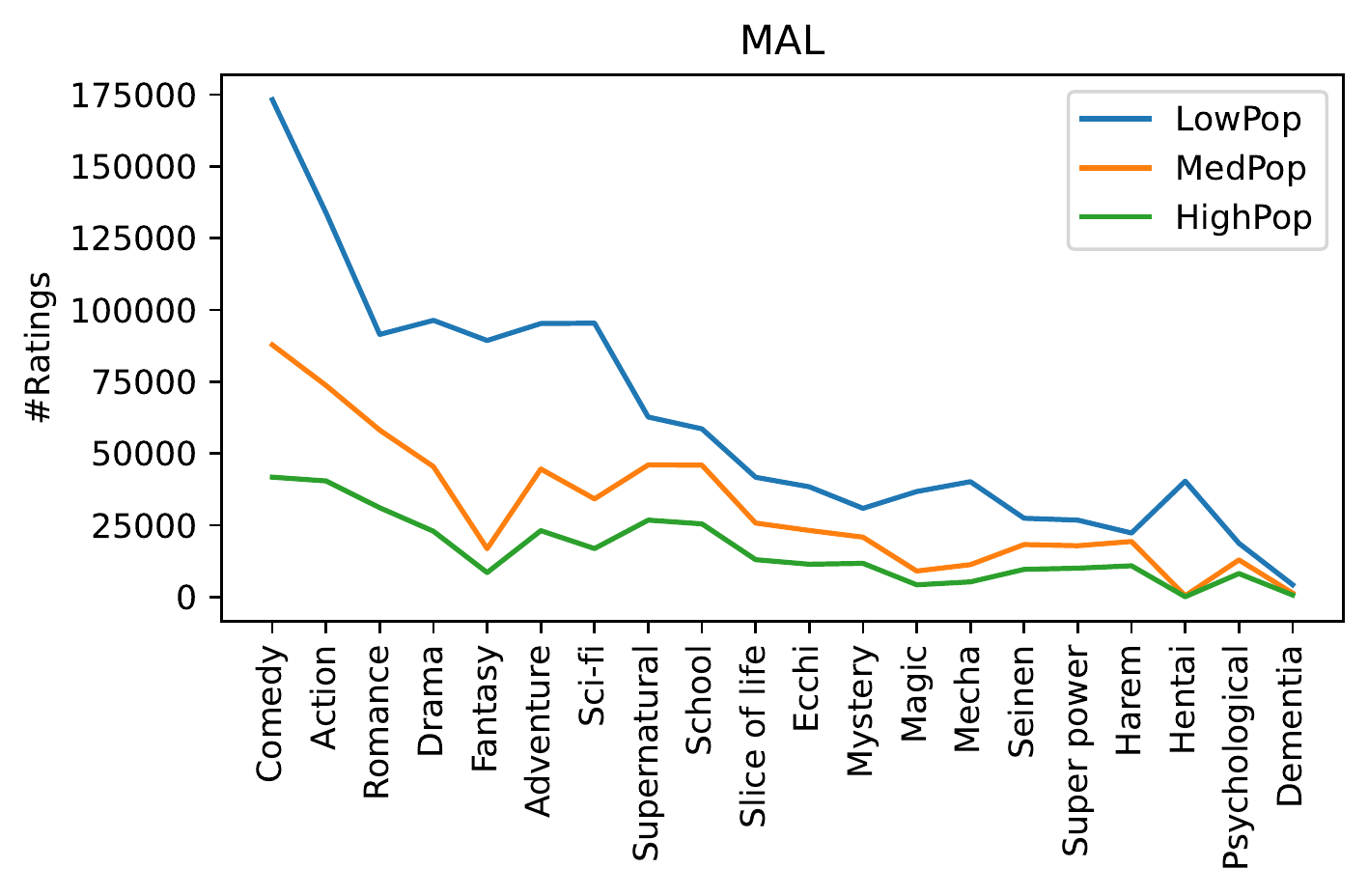}
	\end{subfigure}
    \begin{subfigure}[b]{\textwidth}
    \hspace{1mm}
		\centering
		\includegraphics[width=0.84\textwidth]{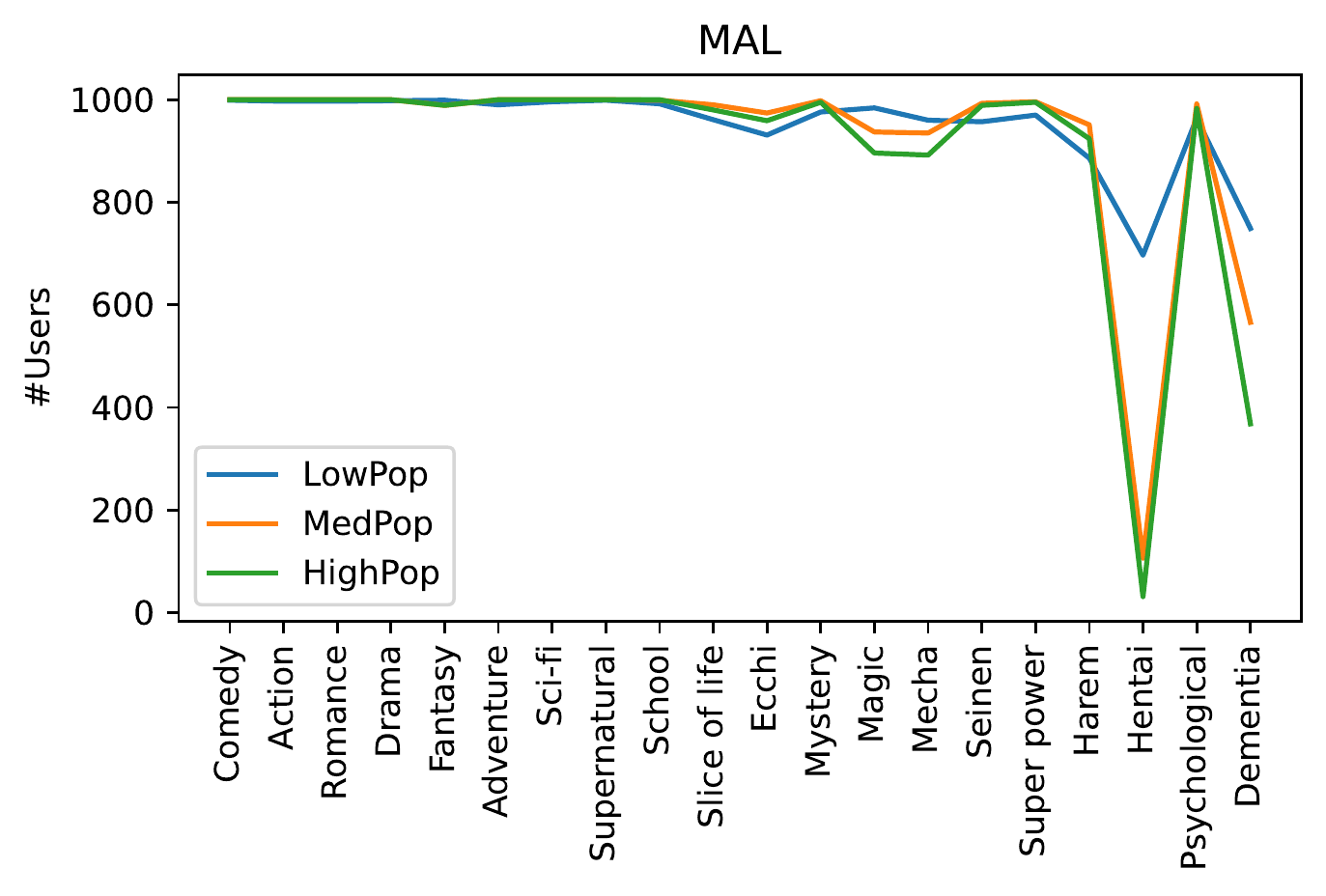}
	\end{subfigure}
	\hfill
 }
	\caption{Genre popularity distribution on the level of ratings (on the left) and on the level of users (on the right) for our three datasets and user groups.} 
	\label{fig:genre_popularities}
\end{figure}

\vspace{2mm} \noindent \textbf{Genre popularity distribution.} 
To get a better understanding of the popularity of the individual genres across the three user groups, in Figure~\ref{fig:genre_popularities}, we plot the genre popularity distribution on the levels of ratings and users. 
The genres are ordered by their overall popularity in terms of ratings across all three user groups, i.e., the most popular genre is the leftmost. On the level of ratings (left plots), we see similar popularity distributions across all user groups. Interestingly, for ML and MAL, the LowPop group has the largest number of ratings across all genres, while for LFM this is the case for the MedPop group. 

On the level of users, we identify similar popularity distributions across all user groups for LFM and ML. However, in the case of MAL, we see a prominent drop for the genre ``Hentai'' when investigating the MedPop and HighPop user groups. This is not the case for the LowPop user group, and thus, the preference for these genres among LowPop users exclusively could lead to an inconsistent recommendation performance for LowPop in the MAL dataset. When relating these results to the rating distributions on the left, we see no drop for the MedPop and HighPop user groups in the case of the ``Hentai'' genre. However, we see an increase in ratings for LowPop for this genre. This again shows the considerable interest of LowPop users for animes associated with the ``Hentai'' genre.

Finally, we also investigated the item popularity distributions across genres and user groups, where we did not inspect any noticeable differences when comparing the user groups on the genre level.

\subsection{Experimental Setup}
Next, we describe the five recommendation algorithms and the evaluation protocol utilized in our study. 

\vspace{2mm} \noindent \textbf{Recommendation algorithms.}
Following our previous research~\cite{kowald2020unfairness,kowald2022ecir}, we formulate the recommendation task as a rating prediction problem by utilizing the Python-based Surprise framework~\cite{hug2020surprise}. Specifically, we use the four collaborative filtering (CF) recommendation algorithms studied in~\cite{kowald2022ecir}. Since our previous work~\cite{kowald2022ecir} also uses the same dataset samples as we do in the present work, we stick to the same hyperparameter settings. Please refer to our source-code shared via GitHub\footnote{\url{https://github.com/domkowald/FairRecSys}} for the exact parameter settings. We refrain from performing any additional hyperparameter optimization since our main interest lies in assessing (relative) differences of our evaluation metrics between the three user groups LowPop, MedPop, and HighPop, and not in comparing a novel recommendation approach to state-of-the-art algorithms. This is also the reason why our focus lies on five traditional and easy understandable recommendation algorithms employed by related work instead of analyzing the performance of recent deep learning architectures, that would also lead to a much higher computational complexity.  

The recommendation algorithms utilized in our study include the two KNN-based algorithms UserKNN and UserKNNAvg, where the latter one incorporates the average rating of the target user and item. We also study Co-Clustering, which is a scalable co-clustering-based CF approach~\cite{george2005scalable}, and NMF, i.e., non-negative matrix factorization~\cite{luo2014efficient}. Additionally, we add a non-CF approach utilized in~\cite{kowald2020unfairness}, namely UserItemAvg, which predicts a baseline estimate using the overall average rating in the dataset and adds preference biases of the target user and item, e.g., if a user tends to give more positive ratings than the average user~\cite{koren2010factor}.

\vspace{2mm} \noindent \textbf{Evaluation protocol.}
Concerning our evaluation protocol, we again follow our previous research~\cite{kowald2020unfairness,kowald2022ecir} and use a random 80/20 train/test split in a 5-fold cross-validation manner. Thus, we train our algorithms on the training set and measure the accuracy of the algorithms on the test set by comparing actual ratings with predicted ratings. By using 5-fold cross-validation, we ensure the robustness of our evaluation protocol, and control for potential fluctuations in the genre proportions or outliers in the recommendation calculations that may be introduced due to the random train/test splits.  

For calculating miscalibration and popularity lift, we use a top-10 recommendation set for the target user, which are the 10 items with the highest predicted rating scores. 
Since our previous research~\cite{kowald2020unfairness,kowald2022ecir} has shown that the LopPop user group typically receives the worst  recommendation accuracy across all user groups, we are especially interested in this user group. 
Therefore, we test for statistical significance using a t-test between LowPop and MedPop as well as between LowPop and HighPop. We report average values across all 5 folds for all metrics and indicate statistical significance only in case it applies for all 5 folds.

\subsection{Evaluation Metrics}
We quantify the inconsistency of recommendation performance using (i) accuracy differences between user groups, (ii) miscalibration, and (iii) popularity lift: 

\vspace{2mm} \noindent \textbf{Accuracy (MAE).} 
We measure accuracy using the well-known mean absolute error (MAE) metric. The MAE of a user $u$ is given by:
\begin{equation}
    MAE(u) = \frac{1}{|R^{test}_u|} \sum_{r_{u, i} \in R^{test}_u} |r_{u, i} - R_{u, i}| 
\end{equation}

Here, the predicted rating score $R_{u, i}$ of user $u$ and item $i$ is compared to the real rating scores $r_{u, i}$ in $u$'s test set $R^{test}_u$. 
We favor MAE over the commonly used root mean squared error (RMSE) metric due to several disadvantages of RMSE, especially regarding the comparison of groups with different numbers of observations (i.e., ratings in our case)~\cite{willmott2005advantages}. We report the MAE of a user group $g$ by averaging the MAE values of all users of $g$. 

To validate our accuracy results in terms of MAE also in top-$n$ recommendation evaluation settings, we also report the well-known Precision and Recall metrics. For this, we classify an item in the test set as relevant if its rating is higher than the average rating in the train set. 

\vspace{2mm} \noindent \textbf{Miscalibration (MC).} 
The calibration metric proposed by Steck~\cite{steck2018} quantifies the similarity of a genre spectrum between user profiles $p$ and actual recommendations $q$. This metric was reinterpreted by Lin et al.~\cite{lin2020} in the form of miscalibration, i.e., the deviation between $p$ and $q$. We follow this definition and calculate the deviation using the Kullback-Leibler (KL) divergence between the distribution of genres in $p$, i.e., $p(c|u)$, and the distribution of genres in $q$, i.e., $q(c|u)$. This is given by:

\begin{align}
KL(p||q) = \sum_{c \in C} p(c|u) \log \frac{p(c|u)}{q(c|u)}
\end{align}

Here, $C$ is the set of all genres in a dataset. Therefore, $KL = 0$ means perfect calibration, and higher $KL$ values (i.e., close to 1) mean miscalibrated recommendations. As in the case of MAE, we report the miscalibration values averaged over all users of a group $g$. 

\vspace{2mm} \noindent \textbf{Popularity lift (PL).} 
The popularity lift metric investigates to what extent recommendation algorithms amplify the popularity bias inherent in the user profiles~\cite{abdollahpouri2019unfairness,abdollahpouri2020connection}. Thus, it quantifies the disproportionate recommendation of more popular items for a given user group $g$ (i.e., LowPop, MedPop, HighPop). We define the group average popularity $GAP_p(g)$ as the average popularity of the items in the user profiles $p$ of group $g$. Similarly, $GAP_q(g)$ is the average popularity of the recommended items for all users of the group $g$. The popularity lift is then given by:

\begin{align}
PL(g) = \frac{GAP_q(g) - GAP_p(g)}{GAP_p(g)}
\end{align}

Here, $PL(g) > 0$ means that the recommendations for $g$ are too popular, $PL(g) < 0$ means that the recommendations for $g$ are not popular enough, and $PL(g) = 0$ would be the ideal value.

\section{Results} 

\begin{table}[t!]
    \centering
    \caption{MAE, MC, and PL results for the LowPop, MedPop, and HighPop user groups. The highest (i.e., worst) results are highlighted in \textbf{bold}. Statistical significance according to a t-test between LowPop and MedPop, and LopPop and HighPop is indicated by * for $p < 0.05$. Rating ranges are shown in brackets. 
    }
    \label{tab:results}
    \resizebox{\textwidth}{!}{
    \begin{tabular}{|l|l|lll|lll|lll|}
    \hline
    & \textbf{\textit{Data}}    & \multicolumn{3}{c|}{\parbox{1.2cm}{\textbf{LFM} [1-1,000]}}     & \multicolumn{3}{c|}{\parbox{1cm}{\textbf{ML} [1-5]}}     & \multicolumn{3}{c|}{\parbox{1cm}{\textbf{MAL} [1-10]}}  \\ \hline
\textbf{\textit{Algorithm}}  & \textbf{\textit{Metric}}   & \textit{MAE} & \textit{MC} & \textit{PL} & \textit{MAE} & \textit{MC} & \textit{PL} & \textit{MAE} & \textit{MC} & \textit{PL} \\ \hline
\multirow{3}{*}{\textbf{UserItemAvg}}& \textit{LowPop}  & \textbf{48.02}* & \textbf{0.52}* & 1.28 & \textbf{0.74}* & \textbf{0.78}* & \textbf{0.70}* & \textbf{0.99}* & \textbf{0.95}* & \textbf{1.12}*\\
& \textit{MedPop}  & 38.48 & 0.48 & \textbf{1.61} & 0.71 & 0.71 & 0.42 & 0.96 & 0.73 & 0.42\\
& \textit{HighPop}  & 45.24 & 0.42 & 1.35 & 0.69 & 0.63 & 0.24 & 0.97 & 0.64 & 0.15\\\hline
\multirow{3}{*}{\textbf{UserKNN}}& \textit{LowPop}  & \textbf{54.32}* & \textbf{0.51}* & 0.52 & \textbf{0.80}* & \textbf{0.75}* & \textbf{0.64}* & \textbf{1.37}* & \textbf{0.92}* & \textbf{0.74}*\\
& \textit{MedPop}  & 46.76 & 0.50 & \textbf{0.82} & 0.75 & 0.69 & 0.37 & 1.34 & 0.72 & 0.22\\
& \textit{HighPop}  & 49.75 & 0.45 & 0.80 & 0.72 & 0.62 & 0.20 & 1.31 & 0.63 & 0.08\\\hline
\multirow{3}{*}{\textbf{UserKNNAvg}}& \textit{LowPop}  & \textbf{50.12}* & \textbf{0.49}* & 0.35 & \textbf{0.76}* & \textbf{0.78}* & \textbf{0.49}* & \textbf{1.00}* & \textbf{0.90}* & \textbf{0.54}*\\
& \textit{MedPop}  & 40.30 & 0.47 & 0.61 & 0.73 & 0.70 & 0.33 & 0.95 & 0.73 & 0.24\\
& \textit{HighPop}  & 46.39 & 0.42 & \textbf{0.64} & 0.70 & 0.61 & 0.20 & 0.95 & 0.64 & 0.11\\\hline
\multirow{3}{*}{\textbf{NMF}}& \textit{LowPop}  & \textbf{42.47}* & \textbf{0.54}* & 0.10 & \textbf{0.75}* & \textbf{0.78}* & \textbf{0.57}* & \textbf{1.01}* & \textbf{0.91}* & \textbf{0.87}*\\
& \textit{MedPop}  & 34.03 & 0.52 & 0.17 & 0.72 & 0.71 & 0.37 & 0.97 & 0.72 & 0.35\\
& \textit{HighPop}  & 41.14 & 0.48 & \textbf{0.33} & 0.70 & 0.63 & 0.22 & 0.95 & 0.63 & 0.13\\\hline
\multirow{3}{*}{\textbf{Co-Clustering}}& \textit{LowPop}  & \textbf{52.60}* & \textbf{0.52}* & 0.68 & \textbf{0.74}* & \textbf{0.77}* & \textbf{0.70}* & \textbf{1.00}* & \textbf{0.90}* & \textbf{1.10}*\\
& \textit{MedPop}  & 40.83 & 0.51 & \textbf{1.04} & 0.71 & 0.70 & 0.43 & 0.96 & 0.72 & 0.42\\
& \textit{HighPop}  & 47.03 & 0.45 & 0.99 & 0.68 & 0.62 & 0.25 & 0.98 & 0.63 & 0.16\\\hline
\end{tabular}
}
\end{table}

In this section, we describe and discuss the results of our study, first on a more general level and then on the level of genres. 

\vspace{2mm} \noindent \textbf{Connection between accuracy, miscalibration and popularity bias.} 
Table~\ref{tab:results} summarizes our results for the three metrics (MAE, MC, PL) over the three user groups (LowPop, MedPop, HighPop), three datasets (LFM, ML, MAL) and five algorithms (UserItemAvg, UserKNN, UserKNNAvg, NFM, Co-Clustering). 
The results presented are averaged over all users and all folds. 
We can see that in the case of ML and MAL, the LowPop user group receive the worst results 
for MAE, MC, and PL. These results are also statistically significant according to a t-test with $p < 0.05$.
For LFM, the LowPop user group also gets the worst results for the MAE and MC metrics. 

However, when looking at the PL metric, we observe different results, namely the highest popularity lift for either MedPop or HighPop. 
This is in line with our previous research~\cite{kowald2020unfairness}, which has shown that the PL metric provides different results for LFM than for ML. One potential difference between music and movies (and also animes) is that music is typically consumed repeatedly (i.e., a user listens to the same artist multiple times), while movies are mostly watched only once. The definition of the PL metric~\cite{lin2020} does not account for repeat consumption patterns~\cite{kotzias2018predicting}, since items are given the same importance regardless of their consumption frequency. This means that items that are consumed for instance 1,000 times by a specific user have the same importance as items that are consumed only once by this user.  

Finally, in Table~\ref{tab:prec_and_rec}, we validate our accuracy results in terms of MAE also in top-$n$ recommendation evaluation settings using the well-known Precision and Recall metrics. To classify relevant items in the test sets, we calculate the average rating in the training sets and treat a test item as relevant if it exceeds this average train rating. We see very similar results as in the case of the MAE metric. This means that in almost all cases, LowPop gets the worst results (i.e., lowest) and HighPop gets the best results (i.e., highest). 

\begin{table}[t!]
    \centering
    \caption{Accuracy results in terms of Precision and Recall. We tested for statistical significance using a t-test between LowPop and MedPop, and LowPop and HighPop users, which is indicated by * for p < 0.05. The best (i.e., highest) results are highlighted in \textbf{bold}. The results are in line with the MAE ones, which means that LowPop receives worst accuracy results, while HighPop receives the best accuracy results.}
    \label{tab:prec_and_rec}
    \resizebox{\textwidth}{!}{
    \begin{tabular}{|l|l|ll|ll|ll|}
    \hline
    & \textbf{\textit{Data}} & \multicolumn{2}{c|}{\textbf{LFM}} & \multicolumn{2}{c|}{\textbf{ML}} & \multicolumn{2}{c|}{\textbf{MAL}}  \\ \hline
\textbf{\textit{Algorithm}}  & \textbf{\textit{Metric}}   & \textit{Precision} & \textit{Recall} & \textit{Precision} & \textit{Recall} & \textit{Precision} & \textit{Recall} \\ \hline
\multirow{3}{*}{\textbf{UserItemAvg}}& \textit{LowPop}  & 0.30 & 0.11 & 0.78* & 0.19* & 0.71* & 0.15* \\
& \textit{MedPop}  & 0.28 & 0.08 & 0.82 & 0.26 & 0.80 & 0.21 \\
& \textit{HighPop}  & \textbf{0.39} & \textbf{0.14} & \textbf{0.83} & \textbf{0.36} & \textbf{0.80} & \textbf{0.33} \\ \hline
\multirow{3}{*}{\textbf{UserKNN}}& \textit{LowPop}  & 0.33* & 0.16 & 0.78* & 0.18* & 0.71* & 0.15* \\
& \textit{MedPop}  & 0.38 & 0.14 & 0.83 & 0.25 & 0.80 & 0.22 \\
& \textit{HighPop}  & \textbf{0.53} & \textbf{0.22} & \textbf{0.83} & \textbf{0.35} & \textbf{0.81} & \textbf{0.34} \\ \hline
\multirow{3}{*}{\textbf{UserKNNAvg}}& \textit{LowPop}  & 0.34 & 0.16 & 0.80* & 0.20* & 0.73* & 0.16 \\
& \textit{MedPop}  & 0.34 & 0.12 & 0.83 & 0.27 & 0.80 & 0.23 \\
& \textit{HighPop}  & \textbf{0.47} & \textbf{0.19} & \textbf{0.83} & \textbf{0.36} & \textbf{0.81} & \textbf{0.36} \\ \hline
\multirow{3}{*}{\textbf{NMF}}& \textit{LowPop}  & 0.34 & 0.16 & 0.70* & 0.14* & 0.67* & 0.13* \\
& \textit{MedPop}  & 0.34 & 0.12 & 0.79 & 0.23 & 0.79 & 0.21 \\
& \textit{HighPop}  & \textbf{0.46} & \textbf{0.19} & \textbf{0.82} & \textbf{0.34} & \textbf{0.81} & \textbf{0.33} \\ \hline
\multirow{3}{*}{\textbf{Co-Clustering}}& \textit{LowPop}  & 0.33 & 0.16* & 0.76* & 0.17* & 0.69* & 0.14* \\
& \textit{MedPop}  & 0.33 & 0.12 & 0.83 & 0.25 & 0.80 & 0.22 \\
& \textit{HighPop}  & \textbf{0.46} & \textbf{0.20} & \textbf{0.84} & \textbf{0.35} & \textbf{0.81} & \textbf{0.34} \\ \hline
\end{tabular}
}
\end{table}

\begin{figure}[!t]
	\centering
\resizebox{\textwidth}{!}{
	\begin{subfigure}[b]{\textwidth}
		\centering
		\includegraphics[width=0.85\textwidth]{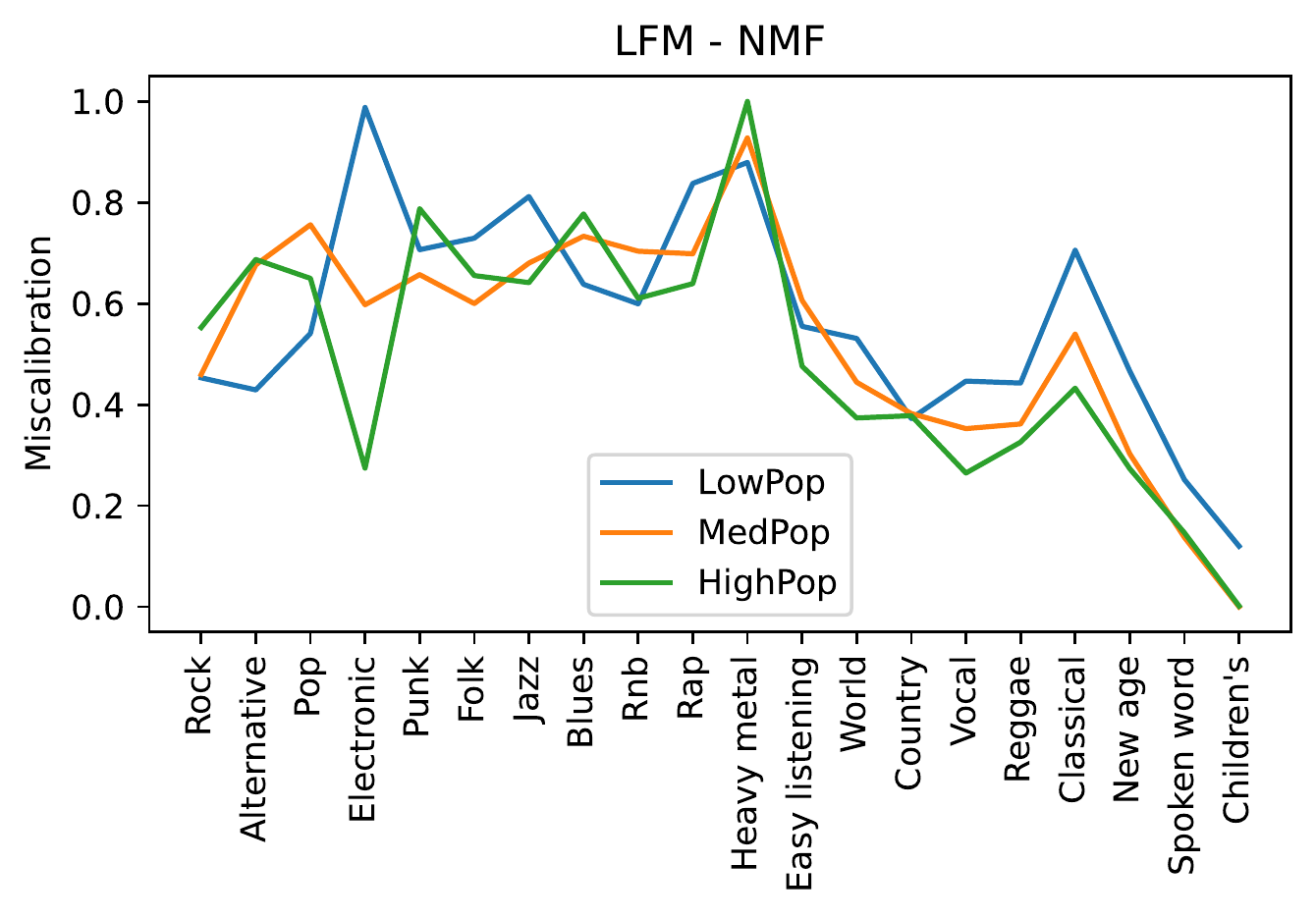}
	\end{subfigure}
	\begin{subfigure}[b]{\textwidth}
 \hspace{-2mm}
		\centering
		\includegraphics[width=0.85\textwidth]{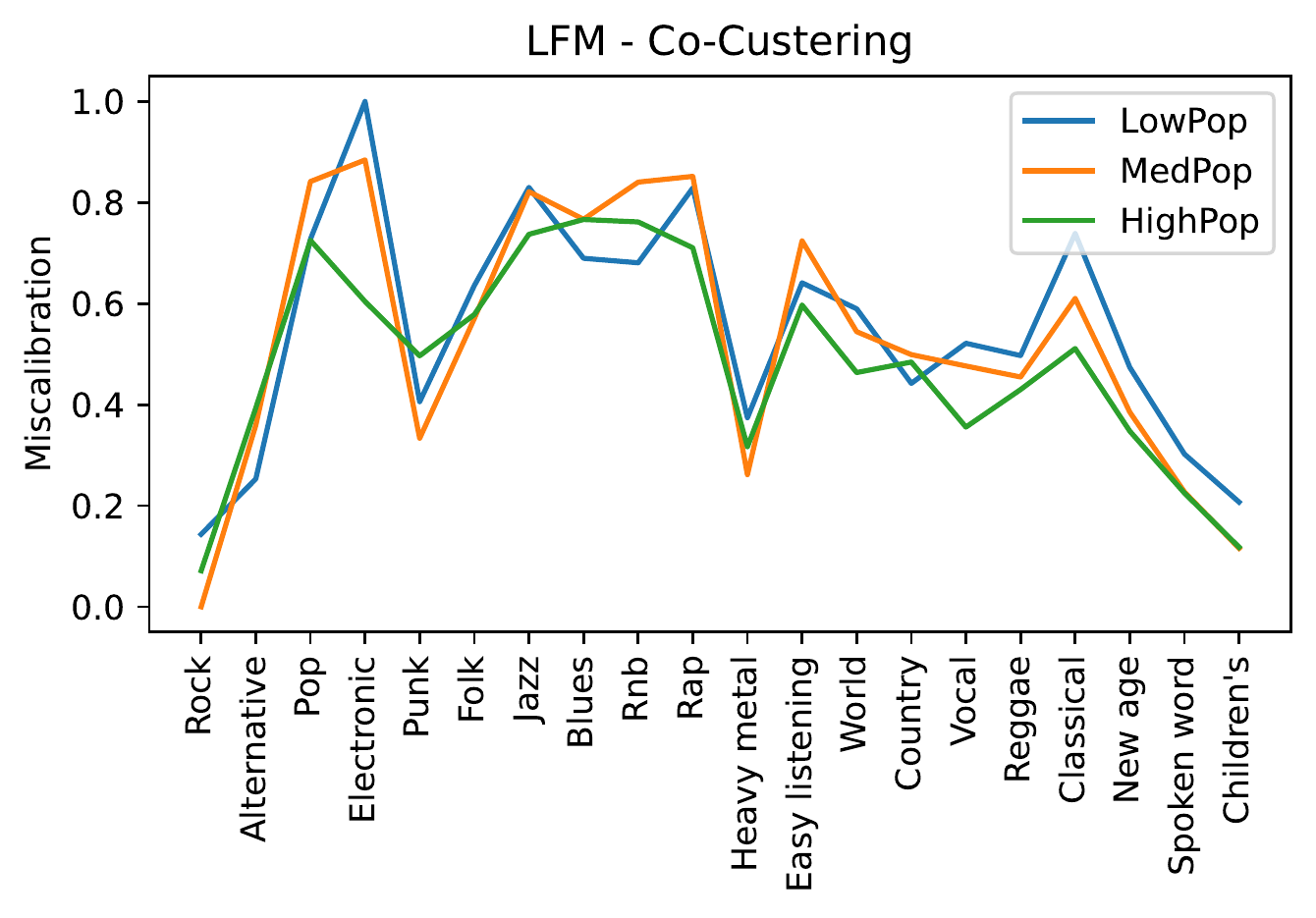}
	\end{subfigure}
	\hfill
 }
\resizebox{\textwidth}{!}{
	\begin{subfigure}[b]{\textwidth}
 \hspace{-2mm}
		\centering
		\includegraphics[width=0.86\textwidth]{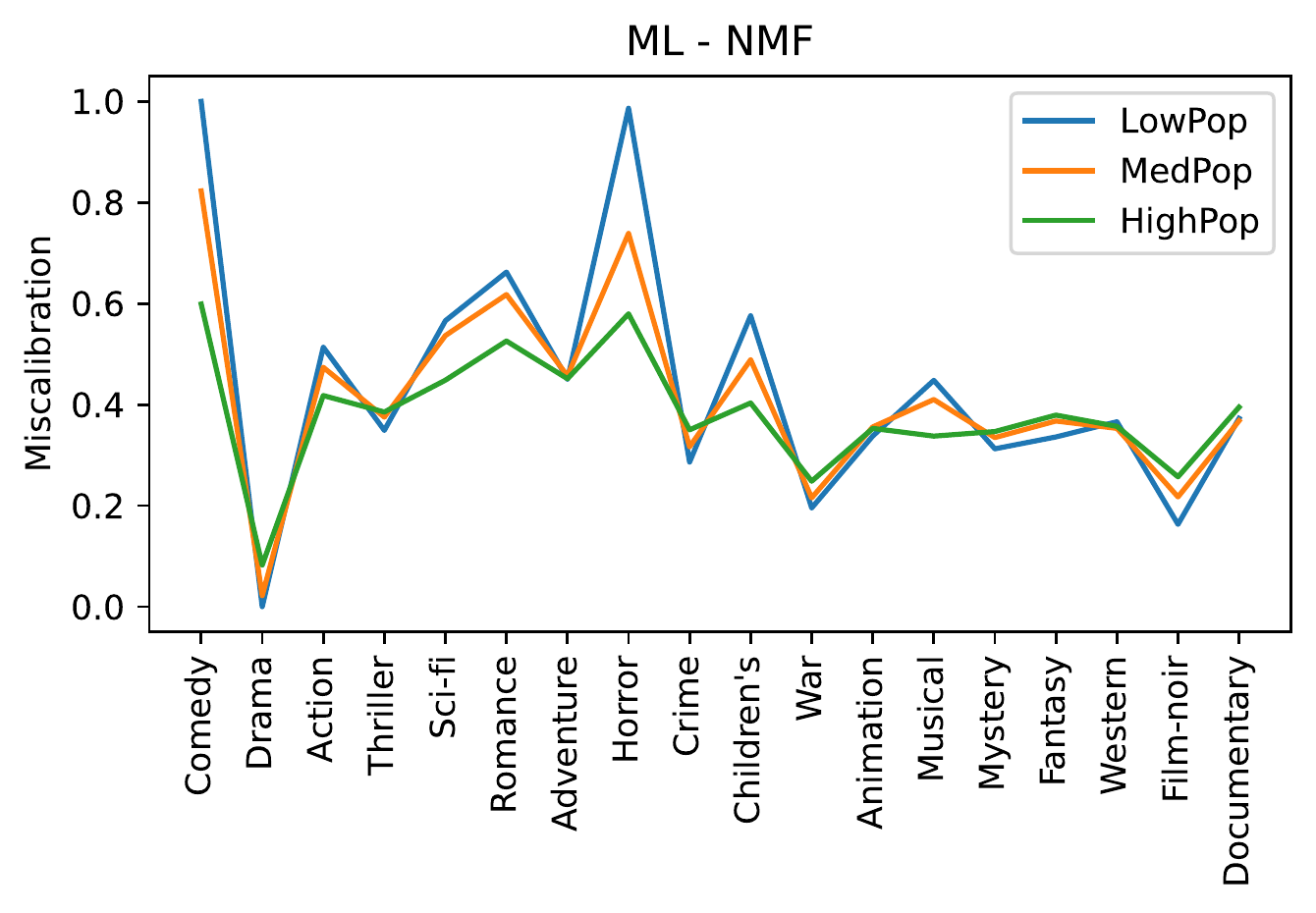}
	\end{subfigure}
	\begin{subfigure}[b]{\textwidth}
		\centering
		\includegraphics[width=0.85\textwidth]{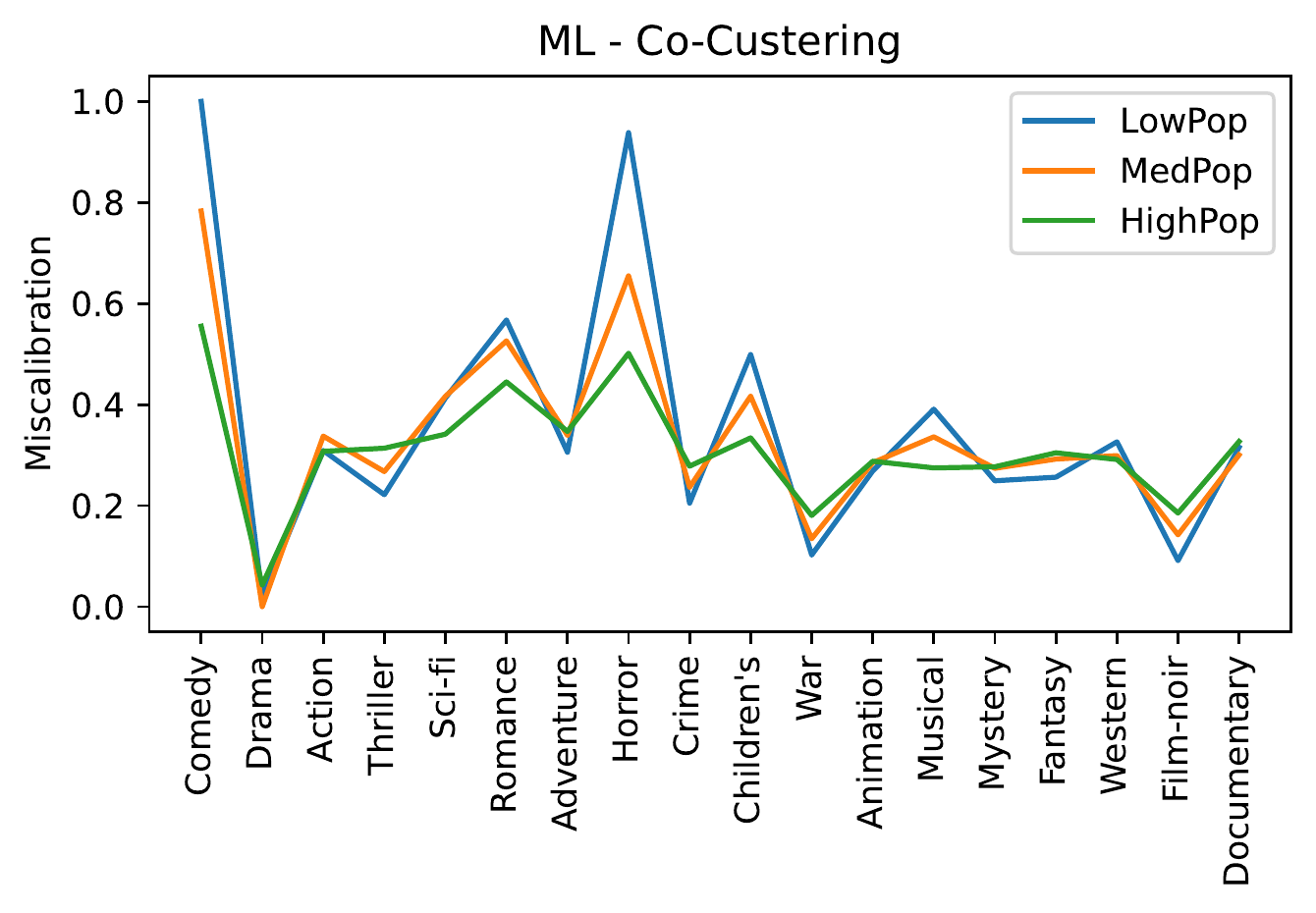}
	\end{subfigure}
	\hfill
 }
 \resizebox{\textwidth}{!}{
	\begin{subfigure}[b]{\textwidth}
		\centering
		\includegraphics[width=0.85\textwidth]{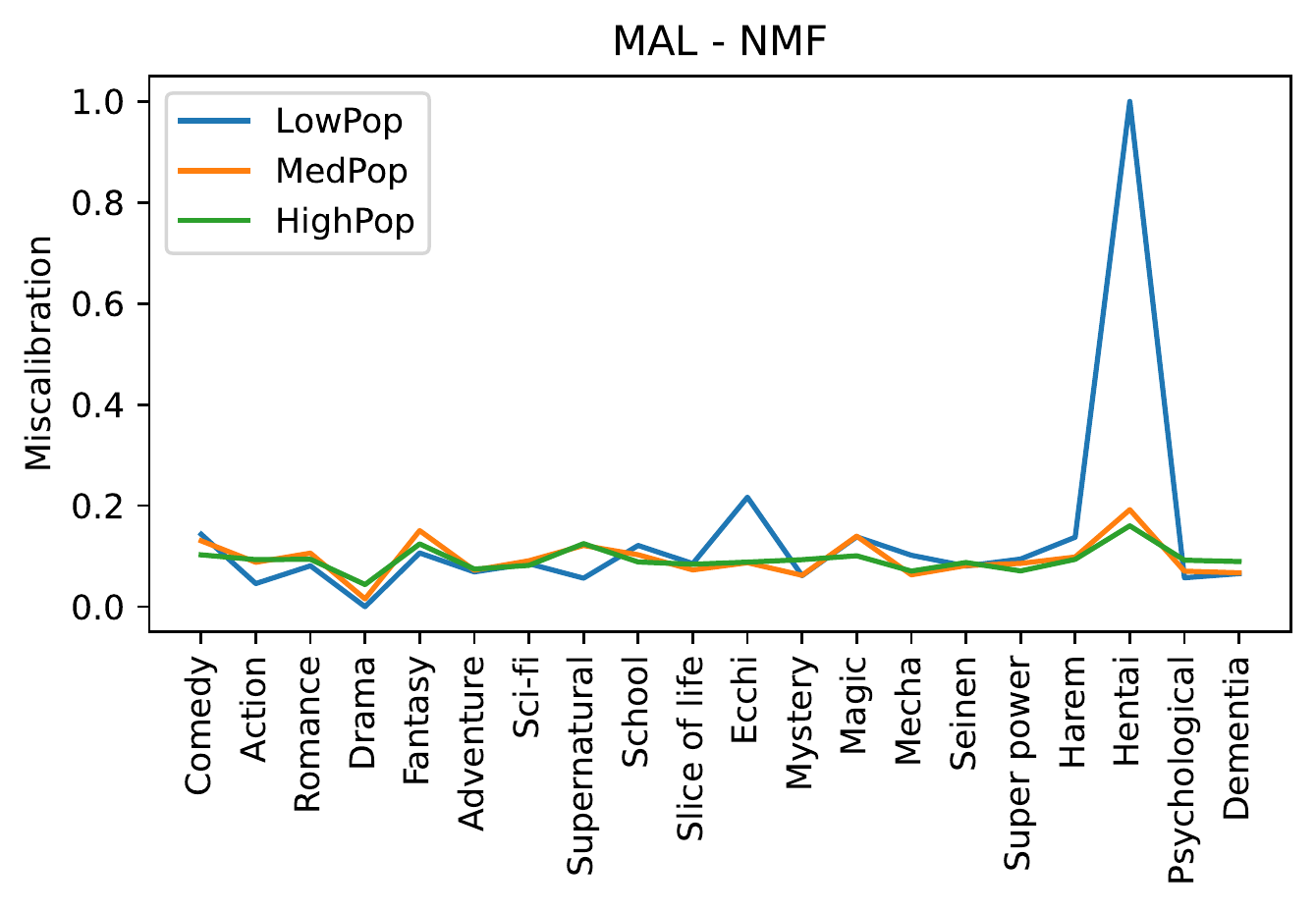}
	\end{subfigure}
    \begin{subfigure}[b]{\textwidth}
    \hspace{1mm}
		\centering
		\includegraphics[width=0.84\textwidth]{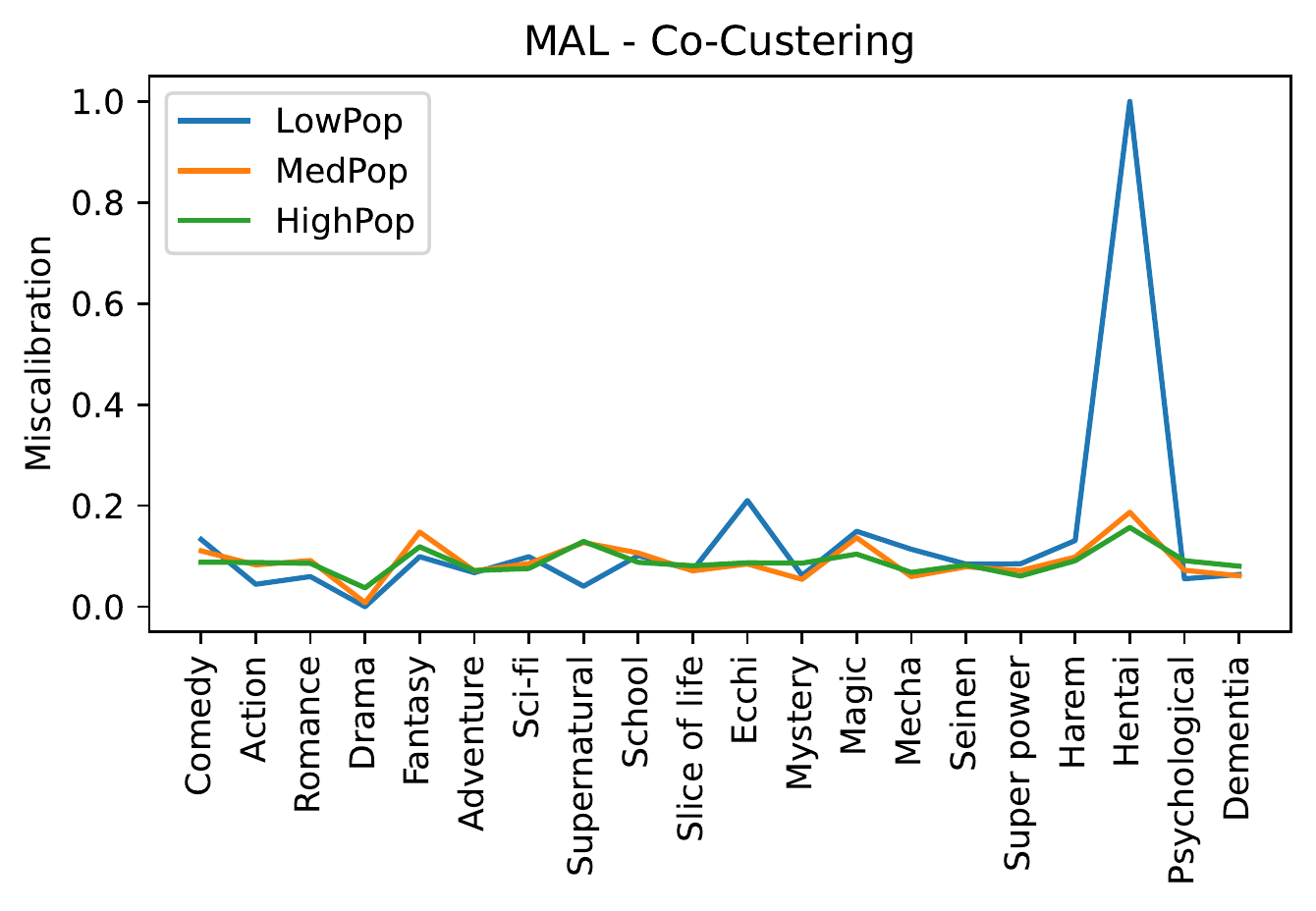}
	\end{subfigure}
	\hfill
 }
	\caption{Influence of different genres on MC for the NMF algorithm (on the left) and Co-Clustering (on the right). We see that some genres highly contribute to inconsistency, especially in case of animes (MAL).
 }
	\label{fig:genre_metric_dist_nmf}
\end{figure}

\vspace{2mm} \noindent \textbf{Influence of genres on inconsistency of recommendations.} 
Furthermore, Figure~\ref{fig:genre_metric_dist_nmf} visualizes the results of our investigation on what genres in the user groups are particularly affecting inconsistency of recommendation performance in terms of miscalibration for the three datasets. We investigate this study for the miscalibration metric only, since we do not observe any particular differences across the genres for the MAE and popularity lift metrics.  
To map the users' miscalibration scores to a genre $g$, we assign the MC score of a user $u$ to all genres listened to $u$. Then for each genre $g$, we calculate the average MC scores of all users of a specific user group who listened to $g$. These values are then plotted in Figure~\ref{fig:genre_metric_dist_nmf} for both the NMF algorithm and the Co-Clustering algorithm. For better readability, we apply min-max normalization in a range of 0 - 1. 

As in the case of Figure~\ref{fig:genre_popularities}, the genres are ordered by their popularity. For the sake of space, we only show the results for NMF and Co-Clustering, which are, in general,  inline with results obtained for the other algorithms. However, our GitHub repository also allows the inspection of the results for the other algorithms. Additionally, for MAL, we exclude 24 genres for which no substantial fluctuations are observed. This leads to 20 shown genres, as in the case of LFM. 

For the MAL dataset and the LowPop group, we observe highly miscalibrated results for the ``Hentai'' genre. In particular, indicated by its position, ``Hentai'' is an unpopular genre for most of the MAL users. However, as also shown in Figure~\ref{fig:genre_popularities}, for users within the LowPop group (and only for this user group), it is a relevant genre that is underrepresented in their recommendation lists. This demonstrates that there are indeed particular genres that contribute to a large extent to recommendation inconsistency for specific user groups. 

\section{Conclusion \& Future Work}
In this paper, we have studied the interconnection between accuracy, miscalibration, and popularity bias for different user groups in three different domains. Here, we measured popularity bias in terms of popularity lift, a metric that compares the popularity of items in recommendation lists to the popularity of items in user profiles. Additionally, we investigated miscalibration, a metric that compares the genre spectrums in user profiles with the ones in recommendation lists. 
We find that, in general, the inconsistency of recommendations in terms of miscalibration and popularity lift is aligned with lower accuracy performance.  

One exception to this is the popularity lift  metric in the case of music recommendations; however, this result is in line with our previous work~\cite{kowald2020unfairness}, in which repeat consumption settings have been studied. Additionally, we find that different genres contribute differently to miscalibration and popularity lift. That finding is particularly pronounced in the case of anime recommendations for LowPop users and for genres that are unpopular among other user groups. 
Another contribution of our work is that we publicly share our datasets and source code investigated in this study with the research community. 

\vspace{2mm} \noindent \textbf{Limitations \& future work.}  
One limitation of our work is that we have focused solely on datasets from the multimedia/entertainment domains, namely music, movies, and animes. Although we have investigated domains with and without repeat consumption patterns, for future work, we plan to also study other domains with respect to accuracy, miscalibration, and popularity bias. This could include recommendations in online marketplaces~\cite{lacic2014towards} or recommendations in social networks~\cite{kowald2013social}. This will contribute to the generalizability of our findings. 
To further strengthen the generalizability of our work, we also plan to conduct further experiments with novel recommendation algorithms employing deep learning-based methods~\cite{lesota2021analyzing}. 

Another limitation of our work is that we have used MAE, Precision, and Recall as the only metrics to measure the accuracy of recommendations. In the future, we plan to extend this by also investigating ranking-based metrics such as nDCG~\cite{jarvelin2002cumulated,lacic2015tackling} as well as metrics that measure the novelty and diversity of recommendations~\cite{castells2022novelty}. 
In this respect, we also plan to enhance our evaluation protocol and move from random train/test splits to temporal train/test splits~\cite{quadrana2018sequence}. 
Finally, we also plan to do experiments with a higher number of user groups with a smaller number of users per group (e.g., 10 groups with 300 users per group). With this, we aim to address a potential limitation with respect to having different popularity tendencies within a group. 

As a general path for future work, we plan to build on the findings of this paper to develop strategies to overcome the inconsistency of recommendation performance across different user groups.   
For example, for particular genres where we find high miscalibration, we aim to research calibration-based debiasing approaches~\cite{abdollahpouri2021user}. 
Another possibility to address popularity bias in recommender system could be to build models based on concepts from psychology~\cite{lex2021psychology}. 
Finally, we plan to investigate novel metrics to measure popularity lift in repeat consumption settings, e.g., music recommendations. Here, we plan to either introduce a weighted variant of the metric or investigate alternative methods for converting implicit feedback (e.g., play counts) into explicit ratings~\cite{pacula2009matrix}.

\vspace{2mm}
\noindent
\textbf{Acknowledgements.} This research was funded by the ``DDAI'' COMET Module within the COMET – Competence Centers for Excellent Technologies Programme, funded by the Austrian Federal Ministry for Transport, Innovation and Technology (bmvit), the Austrian Federal Ministry for Digital and Economic Affairs (bmdw), the Austrian Research Promotion Agency (FFG), the province of Styria (SFG) and partners from industry and academia. Additionally, this work was funded by the Austrian Science Fund (FWF): P33526 and DFH-23. 

\bibliographystyle{splncs04}
\bibliography{references}

\begin{thebibliography}{10}
\providecommand{\url}[1]{\texttt{#1}}
\providecommand{\urlprefix}{URL }
\providecommand{\doi}[1]{https://doi.org/#1}

\bibitem{abdollahpouri2019managing}
Abdollahpouri, H., Burke, R., Mobasher, B.: Managing popularity bias in
  recommender systems with personalized re-ranking. In: The thirty-second
  international flairs conference (2019)

\bibitem{abdollahpouri2019impact}
Abdollahpouri, H., Mansoury, M., Burke, R., Mobasher, B.: The impact of
  popularity bias on fairness and calibration in recommendation. arXiv preprint
  arXiv:1910.05755  (2019)

\bibitem{abdollahpouri2019unfairness}
Abdollahpouri, H., Mansoury, M., Burke, R., Mobasher, B.: The unfairness of
  popularity bias in recommendation. arXiv preprint arXiv:1907.13286  (2019)

\bibitem{abdollahpouri2020connection}
Abdollahpouri, H., Mansoury, M., Burke, R., Mobasher, B.: The connection
  between popularity bias, calibration, and fairness in recommendation. In:
  Fourteenth ACM conference on recommender systems. pp. 726--731 (2020)

\bibitem{abdollahpouri2021user}
Abdollahpouri, H., Mansoury, M., Burke, R., Mobasher, B., Malthouse, E.:
  User-centered evaluation of popularity bias in recommender systems. In:
  Proceedings of the 29th ACM Conference on User Modeling, Adaptation and
  Personalization. pp. 119--129 (2021)

\bibitem{adomavicius2011improving}
Adomavicius, G., Kwon, Y.: Improving aggregate recommendation diversity using
  ranking-based techniques. IEEE Transactions on Knowledge and Data Engineering
   \textbf{24}(5),  896--911 (2011)

\bibitem{baeza2020bias}
Baeza-Yates, R.: Bias in search and recommender systems. In: Fourteenth ACM
  Conference on Recommender Systems. pp.~2--2 (2020)

\bibitem{castells2022novelty}
Castells, P., Hurley, N., Vargas, S.: Novelty and diversity in recommender
  systems. In: Recommender systems handbook, pp. 603--646. Springer (2022)

\bibitem{ekstrand2018all}
Ekstrand, M.D., Tian, M., Azpiazu, I.M., Ekstrand, J.D., Anuyah, O., McNeill,
  D., Pera, M.S.: All the cool kids, how do they fit in?: Popularity and
  demographic biases in recommender evaluation and effectiveness. In:
  Conference on fairness, accountability and transparency. pp. 172--186. PMLR
  (2018)

\bibitem{george2005scalable}
George, T., Merugu, S.: A scalable collaborative filtering framework based on
  co-clustering. In: Fifth IEEE International Conference on Data Mining
  (ICDM'05). pp. 4--pp. IEEE (2005)

\bibitem{harper2015theMD}
Harper, F.M., Konstan, J.A.: The movielens datasets: History and context. ACM
  Trans. Interact. Intell. Syst.  \textbf{5},  19:1--19:19 (2015)

\bibitem{hug2020surprise}
Hug, N.: Surprise: A python library for recommender systems. Journal of Open
  Source Software  \textbf{5}(52), ~2174 (2020)

\bibitem{jarvelin2002cumulated}
J{\"a}rvelin, K., Kek{\"a}l{\"a}inen, J.: Cumulated gain-based evaluation of ir
  techniques. ACM Transactions on Information Systems (TOIS)  \textbf{20}(4),
  422--446 (2002)

\bibitem{koren2010factor}
Koren, Y.: Factor in the neighbors: Scalable and accurate collaborative
  filtering. ACM Transactions on Knowledge Discovery from Data (TKDD)
  \textbf{4}(1),  1--24 (2010)

\bibitem{kotzias2018predicting}
Kotzias, D., Lichman, M., Smyth, P.: Predicting consumption patterns with
  repeated and novel events. IEEE Transactions on Knowledge and Data
  Engineering  \textbf{31}(2),  371--384 (2018)

\bibitem{kowald2013social}
Kowald, D., Dennerlein, S., Theiler, D., Walk, S., Trattner, C.: The social
  semantic server: A framework to provide services on social semantic network
  data. In: 9th International Conference on Semantic Systems, I-SEMANTICS 2013.
  pp. 50--54. CEUR (2013)

\bibitem{kowald2022ecir}
Kowald, D., Lacic, E.: Popularity bias in collaborative filtering-based
  multimedia recommender systems. In: Boratto, L., Faralli, S., Marras, M.,
  Stilo, G. (eds.) Advances in Bias and Fairness in Information Retrieval. pp.
  1--11. Springer International Publishing, Cham (2022)

\bibitem{kowald2021support}
Kowald, D., Muellner, P., Zangerle, E., Bauer, C., Schedl, M., Lex, E.: Support
  the underground: characteristics of beyond-mainstream music listeners. EPJ
  Data Science  \textbf{10}(1), ~14 (2021)

\bibitem{kowald2020unfairness}
Kowald, D., Schedl, M., Lex, E.: The unfairness of popularity bias in music
  recommendation: A reproducibility study. In: European conference on
  information retrieval. pp. 35--42. Springer (2020)

\bibitem{lacic2014towards}
Lacic, E., Kowald, D., Parra, D., Kahr, M., Trattner, C.: Towards a scalable
  social recommender engine for online marketplaces: The case of apache solr.
  In: Proceedings of the 23rd International Conference on World Wide Web. pp.
  817--822 (2014)

\bibitem{lacic2015tackling}
Lacic, E., Kowald, D., Traub, M., Luzhnica, G., Simon, J.P., Lex, E.: Tackling
  cold-start users in recommender systems with indoor positioning systems. In:
  Poster Proceedings of the 9th $\{$ACM$\}$ Conference on Recommender Systems.
  ACM (2015)

\bibitem{lesota2021analyzing}
Lesota, O., Melchiorre, A., Rekabsaz, N., Brandl, S., Kowald, D., Lex, E.,
  Schedl, M.: Analyzing item popularity bias of music recommender systems: are
  different genders equally affected? In: Proceedings of the 15th ACM
  Conference on Recommender Systems. pp. 601--606 (2021)

\bibitem{lex2021psychology}
Lex, E., Kowald, D., Seitlinger, P., Tran, T.N.T., Felfernig, A., Schedl, M.,
  et~al.: Psychology-informed recommender systems. Foundations and
  Trends{\textregistered} in Information Retrieval  \textbf{15}(2),  134--242
  (2021)

\bibitem{lin2020}
Lin, K., Sonboli, N., Mobasher, B., Burke, R.: Calibration in collaborative
  filtering recommender systems: A user-centered analysis. In: Proceedings of
  the 31st ACM Conference on Hypertext and Social Media. p. 197–206. HT '20,
  Association for Computing Machinery, New York, NY, USA (2020)

\bibitem{luo2014efficient}
Luo, X., Zhou, M., Xia, Y., Zhu, Q.: An efficient non-negative
  matrix-factorization-based approach to collaborative filtering for
  recommender systems. IEEE Transactions on Industrial Informatics
  \textbf{10}(2),  1273--1284 (2014)

\bibitem{pacula2009matrix}
Pacula, M.: A matrix factorization algorithm for music recommendation using
  implicit user feedback. Maciej Pacula  (2009)

\bibitem{quadrana2018sequence}
Quadrana, M., Cremonesi, P., Jannach, D.: Sequence-aware recommender systems.
  ACM Computing Surveys (CSUR)  \textbf{51}(4),  1--36 (2018)

\bibitem{schedl2016lfm}
Schedl, M.: The lfm-1b dataset for music retrieval and recommendation. In:
  Proceedings of the 2016 ACM on international conference on multimedia
  retrieval. pp. 103--110 (2016)

\bibitem{steck2018}
Steck, H.: Calibrated recommendations. In: Proceedings of the 12th ACM
  Conference on Recommender Systems. p. 154–162. RecSys '18, Association for
  Computing Machinery, New York, NY, USA (2018)

\bibitem{willmott2005advantages}
Willmott, C.J., Matsuura, K.: Advantages of the mean absolute error (mae) over
  the root mean square error (rmse) in assessing average model performance.
  Climate research  \textbf{30}(1),  79--82 (2005)

\end{thebibliography}

\end{document}